\documentclass[12pt]{article}
\usepackage{jhep-mod}
\usepackage{bm}
\usepackage{amssymb,amsmath,amsthm}
\usepackage{mathrsfs}
\usepackage[utf8]{inputenc}
\usepackage{enumerate}
\usepackage{hyperref}
\hypersetup{colorlinks=true} 
\usepackage{xcolor}
\usepackage{appendix}
\usepackage{graphicx}
\usepackage{dcolumn}
\usepackage{bm}
\usepackage{multirow}
\usepackage{float}
\usepackage{tikz}
\definecolor{purple}{rgb}{1,0,1}
\definecolor{lime}{HTML}{A6CE39} 

\newcommand{\red}[1]{{\color{red} #1}}

\newcommand{\orcidicon}{%
	\begin{tikzpicture}
	\draw[lime, fill=lime] (0,0) 
		circle [radius=0.16] 
		node[white] {{\fontfamily{qag}\selectfont \tiny ID}};
	\draw[white, fill=white] (-0.0625,0.095) 
		circle [radius=0.007];
	\end{tikzpicture}
	\hspace{-3mm}
}
\newcommand\orcidDel{{\href{https://orcid.org/0000-0003-4158-202X}{\orcidicon}}}
\newcommand\orcidMatt{{\href{https://orcid.org/0000-0003-1088-6485}{\orcidicon}}}
\begin{document}
\title{\huge Kochen--Specker theorem revisited}
\author{\Large Del Rajan\orcidDel{}\!\! {\sf and} Matt Visser\orcidMatt{}}
\emailAdd{del.rajan@sms.vuw.ac.nz, delrajan30@gmail.com}
\affiliation{School of Mathematics and Statistics, Victoria University of Wellington, \\
\null\quad PO Box 600, Wellington 6140, New Zealand}
\emailAdd{\break\null\quad\quad\quad\, matt.visser@sms.vuw.ac.nz}
\abstract{

\noindent
The Kochen--Specker theorem is a basic and fundamental 50 year old non-existence result affecting the foundations of quantum mechanix, demonstrating the impossibility of consistently assigning truth values to certain quantum propositions, thereby strongly implying the lack of any meaningful notion of ``quantum realism''. Indeed the Kochen--Specker theorem is typically interpreted in terms of requiring ``contextuality'' (roughly speaking, context-dependent ``reality'') in quantum physics. 
The original proofs of the Kochen--Specker theorem proceeded via finding brute force counter-examples; often quite complicated and subtle (albeit mathematically ``elementary'')  counter-examples. Only more recently have somewhat more ``geometrical'' proofs been developed. We present herein yet another simplified geometrical proof of the Kochen--Specker theorem, one that is valid for any number of dimensions, that minimizes the technical machinery involved, and makes the seriousness of the issues raised manifest.

\bigskip
\noindent
{\sc Pacs:} 03.65.-w; 03.65.Aa; 03.65.Ta


\bigskip
\noindent
{\sc arXiv:}  1708.01380 [quant-ph]

\bigskip
\noindent
{\sc Dated:} 10 April 2019; \LaTeX-ed \today

\bigskip
\noindent
\red{
{\sc Warning:} \\
An earlier version of this document (August 2017) contained a number of serious measure-theoretic flaws. 
We believe we have now fixed all of those earlier problems.
}
}

\maketitle
\theoremstyle{definition}
\newtheorem{lemma}{Lemma}
\parindent0pt
\parskip7pt
\def\C{\mathcal{C}}
\def\tr{{\mathrm{tr}}}
\def\d{{\mathrm{d}}}
\def\g{{\mathfrak{g}}}
\def\dbar{{\mathchar'26\mkern-12mu \d}} 
\def\Hilbert{{\mathcal{H}}}
\def\H{{\mathrm{H}}}
\def\R{{\mathrm{R}}}
\def\E{{\mathrm{E}}}
\def\nn{\nonumber}
\enlargethispage{20pt}
\vspace{-15pt}
\section{Introduction}
The Kochen--Specker theorem, (sometimes called the Bell--Kochen--Specker theorem), 
demonstrates the impossibility of consistently assigning $\{0,1\}$ truth values to quantum propositions.
We shall phrase our discussion in terms of real Hilbert spaces, noting that a complex Hilbert space can always be viewed as a real Hilbert space of double the dimensionality, $\mathbb{C}^n \sim \mathbb{R}^{2n}$.
The Kochen--Specker theorem was originally proved some 50 years ago by explicitly finding a set of 117 distinct projection operators on 3-dimensional Hilbert space~\cite{Kochen:1967,Bell:1966}, and then showing that there was no way to consistently assign values in $\{0,1\}$ to these projection operators. (That is, these 117 ``quantum questions'' that one might ask could not be consistently assigned yes-no answers.) A later version of the Kochen--Specker theorem reduced the number of projection operators to 33~\cite{Peres}. This was further reduced to 24~\cite{Peres}, to 20~\cite{Kernaghan}, and then  to 18~\cite{Cabello1,Cabello2}, at the cost of slightly increasing the dimension of the Hilbert space to 4. 
Ultimately the number of projection operators was  further reduced to 13 in an 8-dimensional Hilbert space in reference~\cite{Peres2}.
Interest in these foundational issues has continued unabated~\cite{Mermin1,Mermin2}, 
with at least two  ``geometrical'' proofs dating from late 1990s that avoid explicit construction of sets of projection operators~\cite{Gill,Calude}. 
Herein  we shall provide yet another even more simplified ``geometrical'' proof of the Kochen--Specker theorem, which, while it is still non-constructive, (proceeding by establishing an inconsistency), is utterly minimal in its technical requirements, and so hopefully instructive.

\section{Statement of the Kochen--Specker theorem}

\vspace{-7pt}
An explicit statement of the Kochen--Specker theorem, (based on the discussion in the Stanford encyclopaedia of philosophy), runs thus:

{\bf Theorem} ({\sf Kochen-Specker --- mathematical version}){\bf:}
\emph{Let $H$ be a Hilbert space of QM state vectors of real dimension $d \geq  3$. Then there is a set $M$ of observables on $H$, containing $n$ elements, such that the following two assumptions are contradictory:
\vspace{-12pt}
\begin{description}
\itemsep-3pt
\item[KS1:]  All $n$ members of $M$ simultaneously have values, that is, they are unambiguously mapped onto real numbers (designated, for specific observables $A, B, C$, ..., by values $v(A)$, $v(B)$, $v(C)$, ...).
\item[KS2:]  Values of observables conform to the following constraints:
\vspace{-5pt}
\begin{description}
\itemsep-3pt
\item[(a)] If $A, B, C$ are all compatible and $C = A+B$, then $v(C) = v(A)+v(B)$.
\item[(b)] If  $A, B, C$ are all compatible and $C = AB$, then $v(C) = v(A)v(B)$.
\item[(c)] $\exists$ at least one observable $X$ with $v(X)\neq 0$.\\
(Here ``compatible'' means that the observables commute.)
\rightline{$\Box$}
\end{description}
\end{description}
\vspace{-7pt}
}
There are several technical issues with this presentation. Without condition {\bf KS2c} the theorem is actually false --- the trivial valuation where for all observables $X$ one sets $v(X)= 0$ provides an explicit counter-example. 
Without condition {\bf KS2c}, $v(I) = v(I^2) = v(I)^2$ only implies $v(I)\in\{0,1\}$.
With condition {\bf KS2c} we have the stronger statement that  $v(X) = v(IX) = v(I)v(X)$, which since $v(X)\neq 0$  implies $v(I)=1$.

A more subtle issue is this: 
Physically, we would like to have $v(zA)=z\, v(A)$, for any $z\in \mathbb{C}$.
But using the conditions  {\bf KS2a} and  {\bf KS2b}  we could only deduce this for rational numbers. 
Extending this to the complex numbers requires us to first construct the real numbers ``on the fly'' using Dedekind cuts, and then to formally construct the complex numbers as an algebraic extension of the field of real numbers --- while this is certainly possible, in a physics context it is rather pointless --- it would seem more reasonable to start with the complex numbers as being given, even if you then need slightly stronger axioms.

{\bf Improved KS2 axioms:}
\vspace{-7pt}
\begin{description}
\itemsep-3pt
\item[(a)] If $[A,B]=0$ and $a,b\in \mathbb{C}$, then $v(aA+bB) = a\,v(A)+b\,v(B)$.
\item[(b)] If $[A,B]=0$ then $v(AB) = v(A)\,v(B)$.
\item[(c)] $\exists$ at least one observable $X$ with $v(X)\neq 0$.
\end{description}
\vspace{-7pt}

\enlargethispage{40pt}
If one accepts these improved {\bf KS2} axioms then immediately
\begin{equation}
v(I)=1; \qquad v(aI)=a; 
\end{equation}
and for any analytic function with a non-zero radius of convergence
\begin{equation}
v(f(A))= f(v(A)).
\end{equation}
\smallskip

\clearpage
Note that this last condition, $v(f(A))= f(v(A))$, is where physics discussions of the Kochen--Specker theorem often \emph{start}. 
Indeed let us write $A=\sum_i  a_i P_i$ where the $a_i$ are real and the $P_i$ are projection operators onto 1-dimensional subspaces; so the projection operators $P_i = |\psi_i\rangle\,\langle\psi_i|$ can be identified with the vectors $|\psi_i\rangle$ which form a basis for the Hilbert space. Then
\begin{equation}
v(A) = v\left(\sum_i  a_i P_i\right) = \sum_i  a_i \; v(P_i).
\end{equation}
This now focusses attention on the valuations $v(P_i)$. Since $P_i^2=P_i$, condition {\bf KS2b} implies that $v(P_i)\in\{0,1\}$; the valuation must be a yes-no valuation. Now consider the identity operator $I = \sum_i P_i$ and note
\begin{equation}
\sum_i v(P_i) = v(I) = 1.
\end{equation}
It is customary to identify the projectors $P_i = |n_i\rangle\, \langle n_i | $ with the corresponding unit vectors $n_i$, (defined up to a sign), with the $n_i$ forming a basis for Hilbert space, and in $d$ dimensions write
\begin{equation}
\sum_{i=1}^d v(n_i)  = 1;    \qquad \qquad v(n_i)  \in\{0,1\}; \qquad\quad v(-n)=v(n). 
\end{equation}
It is the \emph{claimed existence} of this function $v(n)$, having the properties stated above for \emph{any arbitrary} basis of Hilbert space, which is the central point of the {\bf KS1} and {\bf KS2} conditions.  This discussion allows us to rephrase the Kochen--Specker theorem in terms of the \emph{non-existence} of such a valuation.

\enlargethispage{10pt}
{\bf Theorem:} ({\sf Kochen-Specker --- physics-based  version}){\bf:}
\emph{For $d\geq3$ there is no valuation $v(n)  : S^{d-1} \to \{0, 1\}$, where $S^{d-1}$
is the unit hypersphere, such that $v(-n) = v(n)$ for all $n$ and
\begin{equation}
\sum_{i=1}^d v(n_i)  = 1,
\end{equation}
for every basis (frame, $d$-bein) of orthogonal unit vectors  $n_i$.
}

It is this statement about bases in Hilbert space that is often more practical to work with, rather than the formulation at the start of this section --- of course without that initial formulation it would be less than clear why the basis formulation is physically interesting.

\section{Yet another proof of the Kochen--Specker theorem}

We will start by looking in a non-traditional place, by considering one-dimensional and two-dimensional Hilbert spaces, before dealing with three-dimensional Hilbert space, (which then settles things for any higher dimensionality).
Since one is trying to prove an inconsistency result, there will be an infinite number of ways of doing so; the question is whether one learns anything new by coming up with a different proof. We shall do so with a modified and simplified ``descent'' argument, one that requires only two steps in the descent process.\footnote{\red{An earlier version of this paper tried to work with notions of the average of the valuation $\overline{v(n)}$ over great circles and hyperspheres. This requires either assuming or deriving some notion of measurability for the sets $v^{-1}(0)$ and $v^{-1}(1)$,  and is much more delicate than we had originally envisaged.  In this version of the article we eschew measurability issues in favour of a simple two-step descent process.}}

\subsection{One dimension}

There is no Kochen--Specker no-go result in one dimension, since in one dimension all operators are multiples of the identity, $A= a I$, and then
\begin{equation}
v(f(A)) = v(f(aI)) = v(f(a)I) = f(a)\, v(I) = f(a). 
\end{equation}
In particular, as long as $f(a)\not\equiv 0$,  (which is implied by the {\bf KS2c} axiom),  then for the (unique) normalized basis vector $n$ we have $v(n) = 1$.

Conversely if we are considering a one-dimensional subspace of a higher-dimensional Hilbert space then the {\bf KS2c} axiom tells us nothing;  for the (unique) normalized basis vector we merely have $v(n)\in\{0,1\}$, and we have no further constraint on the valuation.

\subsection{Two dimensions}

There is no Kochen--Specker no-go result in two dimensions, but there are still quite interesting things to say.
Consider the valuation $v : S^1 \to \{0, 1\}$ (where $S^1$
is the unit circle) such that $v(-n) = v(n)$ for all $n$ and
\begin{equation}
v(n_1) + v(n_2) = 1 
\end{equation}
for every dyad (every pair of orthogonal unit vectors) $n_1$, $n_2$.
Indeed in two dimensions we \emph{can} construct such a valuation. Re-characterize $n_1$ and $n_2$ in terms of the angle they make with (say) the $x$ axis; then the constraints we want to impose are
\begin{equation}
v(\theta)=v(\theta+\pi); \qquad\qquad v(\theta) + v(\theta\pm\textstyle{\pi\over2}) = 1.
\end{equation}
But these conditions are easily solved: Let $g(\theta)$ be some arbitrary (not necessarily continuous) function mapping the interval $\left[0,\textstyle{\pi\over2}\right) \to \{0,1\}$,  and define
\begin{equation}
v(\theta) =  \left\{  \begin{array}{cl} 
\vphantom{\Big |} g(\theta) & \hbox{ for } \theta \in  [0,\textstyle{\pi\over2});\\
\vphantom{\Big |} 1-g(\theta- \textstyle{\pi\over2}) & \hbox{ for } \theta \in  [\textstyle{\pi\over2},\pi);\\
\vphantom{\Big |} g(\theta-\pi) & \hbox{ for }\theta \in  [\pi,\textstyle{3\pi\over2});\\
\vphantom{\Big |} 1- g(\theta - \textstyle{3\pi\over2}) & \hbox{ for }\theta \in  [\textstyle{3\pi\over2},2\pi).\\
   \end{array}\right. 
\end{equation}
So the existence of a Kochen--Specker valuation is easily verified in two dimensions, and because points separated by $\pi/2$ radians must be given opposite valuations, the image $v(S^1)$ is automatically 50\%--50\% zero-one.  
Note in particular that the function $v(\theta)$ \emph{cannot} be everywhere continuous. 
(We will recycle these results repeatedly when we turn to three and higher dimensions.)

\subsection{Three dimensions}

It is in 3 dimensions that things first get interesting. We are interested in  valuations $v : S^2 \to \{0, 1\}$, (where $S^2$
is the unit 2-sphere), such that $v(-n) = v(n)$ for all $n$ and
\begin{equation}
v(n_1) + v(n_2) + v(n_3) = 1 
\end{equation}
for every triad (every triplet of orthogonal unit vectors) $n_1$, $n_2$, $n_3$.
In the argument below we shall make extensive use of the great circles $S^1$ in the unit 2-sphere $S^2$.

{\bf Lemma:}
On any great circle in $S^2$, under the conditions given above, the valuation is either  50\%--50\% zero-one (as in two dimensions), or is 100\% zero (identically zero). \hfill $\Box$

{\bf Proof:}\\
Pick any great circle and for convenience align it with the equator. \\
Now look at the poles:
\vspace{-15pt}
\begin{itemize}
\itemsep-3pt
\item 
If $v(\hbox{poles})=1$, then $v(\hbox{equator})\equiv 0$ is identically zero.\\
(Since points on the equator will be part of some triad that includes the unit vector pointing to the poles.)
\item 
If $v(\hbox{poles})=0$, then any dyad lying in the equator will satisfy the conditions of the two dimensional argument given above, and so will be 50\%--50\% zero-one.
\end{itemize}
\vspace{-10pt}
\hfill $\Box$

Now bootstrap this to a modified ``great circle descent'' argument,  one that needs only two steps in the descent process.
We start with a purely geometrical result.

\begin{figure}[!htbp]
\begin{center}
\includegraphics[scale=1.00]{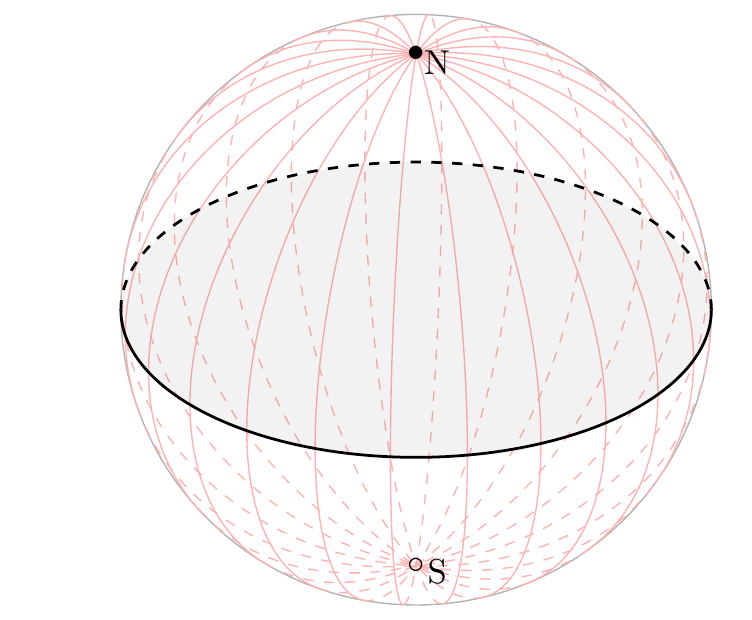}
\caption{Setup with $v(\hbox{poles})=1$, $v(\hbox{equator})\equiv 0$, and $v(\hbox{meridians})$ 50\%--50\% zero-one.}
\label{F:longitudes}
\end{center}
\end{figure}

From the argument above if we arrange $v(\hbox{poles})=1$, then $v(\hbox{equator})\equiv 0$, and for each line of longitude $v(\hbox{meridian})$ will be 50\%--50\% zero-one. (See figure \ref{F:longitudes}.)

\begin{figure}[!htbp]
\begin{center}
\includegraphics[scale=1.00]{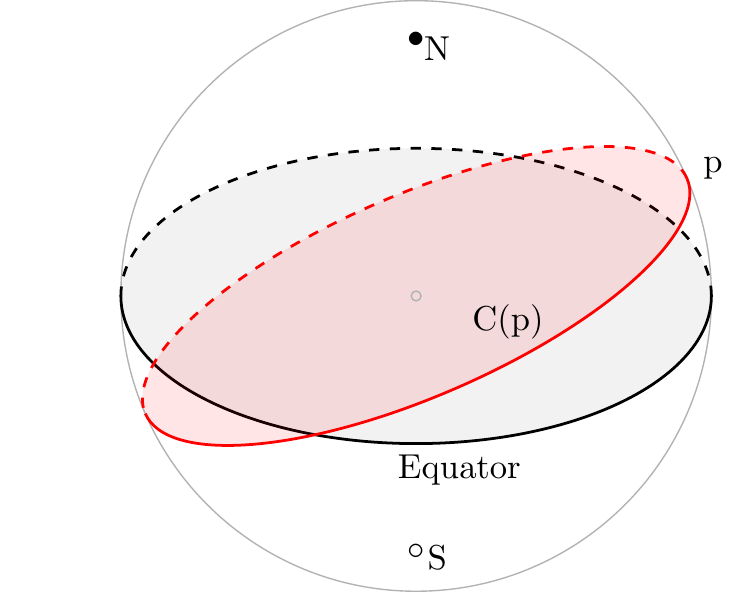}
\caption{Descent great circle $C(p)$, with northernmost point at $p$.}
\label{F:descent-1}
\end{center}
\end{figure}

\begin{figure}[!htbp]
\begin{center}
\includegraphics[scale=0.50]{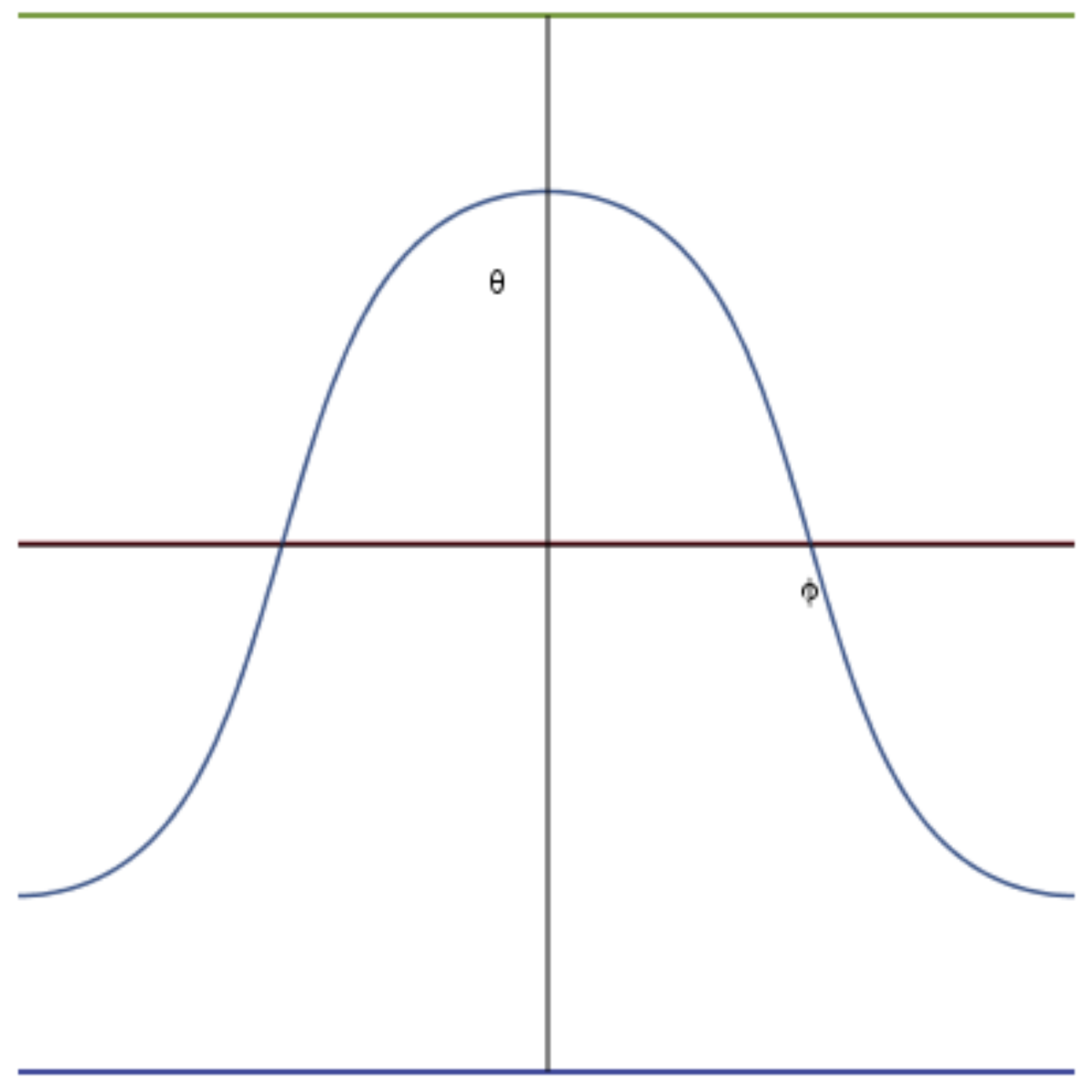}
\caption{Descent great circle represented in terms of $\theta(\phi)$.}
\label{F:descent-2}
\end{center}
\end{figure}

{\bf Definition:}  A ``great circle descent''  $C(p)$ through a point $p$ on the sphere is a great circle that starts off at constant latitude.
(So the point $p$ is either the northernmost or southernmost point on the great circle.  See figures~\ref{F:descent-1} and~\ref{F:descent-2}.)\hfill $\Box$

\newpage
{\bf Lemma:} Let $q$ be any other point at the same longitude as $p$ (the same meridian) that is closer to the equator than $p$. Then there exists a point $r$ such that $r$ lies on the great circle descent through $p$, and $q$ lies on the great circle descent through $r$. \hfill $\Box$

\begin{figure}[!htbp]
\begin{center}
\includegraphics[scale=0.750]{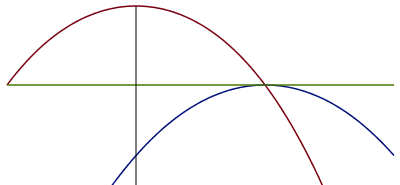}
\caption{Example of a two-step descent with net motion along a meridian.}
\label{F:descent-2-step}
\end{center}
\end{figure}

That is $r\in C(p)$ and $q\in C(r)$, so one can always zig-zag directly to towards the equator via exactly two great circle descents.
Note that this is a much easier geometric result than that used in the Gill--Keane~\cite{Gill} or Calude--Hertling---Svozil~\cite{Calude} approaches where a finite but possibly large number of great circle descents is used to get to any point closer to the equator, not necessarily at the same longitude.
(See figure~\ref{F:descent-2-step}.)

{\bf Proof:} Using spherical coordinates $(\theta,\phi)$ let the generic point $x$ be represented by the 3-vector 
\begin{equation} 
\vec x = (\cos\theta \cos\phi, \cos\theta \sin\phi,\sin\theta).
\end{equation} 
(Somewhat unusually, we adopt conventions close to the usual latitude nomenclature:  $\theta=+\pi/2$ represents the north pole, $\theta=0$ represents the equator, while while $\theta=-\pi/2$ represents the south pole. Doing this simplifies some of the formulae below.)\\
Now let the specific point $p$ of interest be represented by the 3-vector 
\begin{equation} 
\vec p = (\cos\theta_p \cos\phi_p, \cos\theta_p \sin\phi_p,\sin\theta_p).
\end{equation} 
Consider the great circle descent $C(p)$. This great circle will be orthogonal to the vector
\begin{equation} 
\vec p_\perp = (\sin\theta_p \cos\phi_p, \sin\theta_p \sin\phi_p,-\cos\theta_p).
\end{equation} 
The entire great circle $C(p)$ will be characterized by $\vec p_\perp \cdot \hat x(\theta,\phi) =0$, that is
\begin{equation} 
\sin\theta_p\cos\theta (\cos\phi_p \cos\phi + \sin\phi_p \sin \phi_p) - \cos\theta_p \sin\theta  = 0,
\end{equation} 
implying
\begin{equation} 
\sin\theta_p\,\cos\theta\, \cos(\phi - \phi_p) = \cos\theta_p \sin\theta.
\end{equation} 
That is
\begin{equation} 
\tan \theta = {\tan \theta_p \cos(\phi-\phi_p)},
\end{equation} 
or more explicitly 
\begin{equation} 
\theta(\phi) = \tan^{-1} \left({\tan \theta_p\; \cos(\phi-\phi_p)}\right).
\end{equation} 
This explicitly yields $\theta(\phi)$ along the entire descent circle $C(p)$. (See figure~\ref{F:descent-2}.)

Note that this descent circle crosses the equator at $\theta=0$, implying $(\phi-\phi_p)=\pm\pi/2$. This occurs at the points $s$ such that $\vec s=\pm(-\sin\phi_p,\cos\phi_p,0)$.

In particular, for the three points $p$, $r$, $q$, (and using $\phi_p=\phi_q$ because we want $p$ and $q$ to have the same longitude), we have
\begin{equation} 
\tan \theta_r = {\tan \theta_p\; \cos(\phi_r-\phi_p)}; \qquad \tan \theta_q = {\tan \theta_r\; \cos(\phi_r-\phi_p)};
\end{equation} 
implying
\begin{equation} 
\tan \theta_q = {\tan \theta_p\; \cos^2(\phi_r-\phi_p)}.
\end{equation} 
That is
\begin{equation} 
 \cos^2(\phi_r-\phi_p) = {\tan \theta_q\over \tan\theta_p}.
\end{equation} 
Alternatively
\begin{equation}
|\phi_r-\phi_p| = \cos^{-1} \sqrt {\tan \theta_q\over \tan\theta_p}.
\end{equation}
The azimuthal difference $|\phi_r-\phi_p|$ tells you exactly how much you have to zig-zag along the descent circles for the net motion to be directly along the line of longitude towards the equator. Note $|\phi_r-\phi_p|$ is real only if you move towards (rather than away from) the equator.
\hfill $\Box$

\begin{figure}[!htbp]
\begin{center}
\includegraphics[scale=0.5]{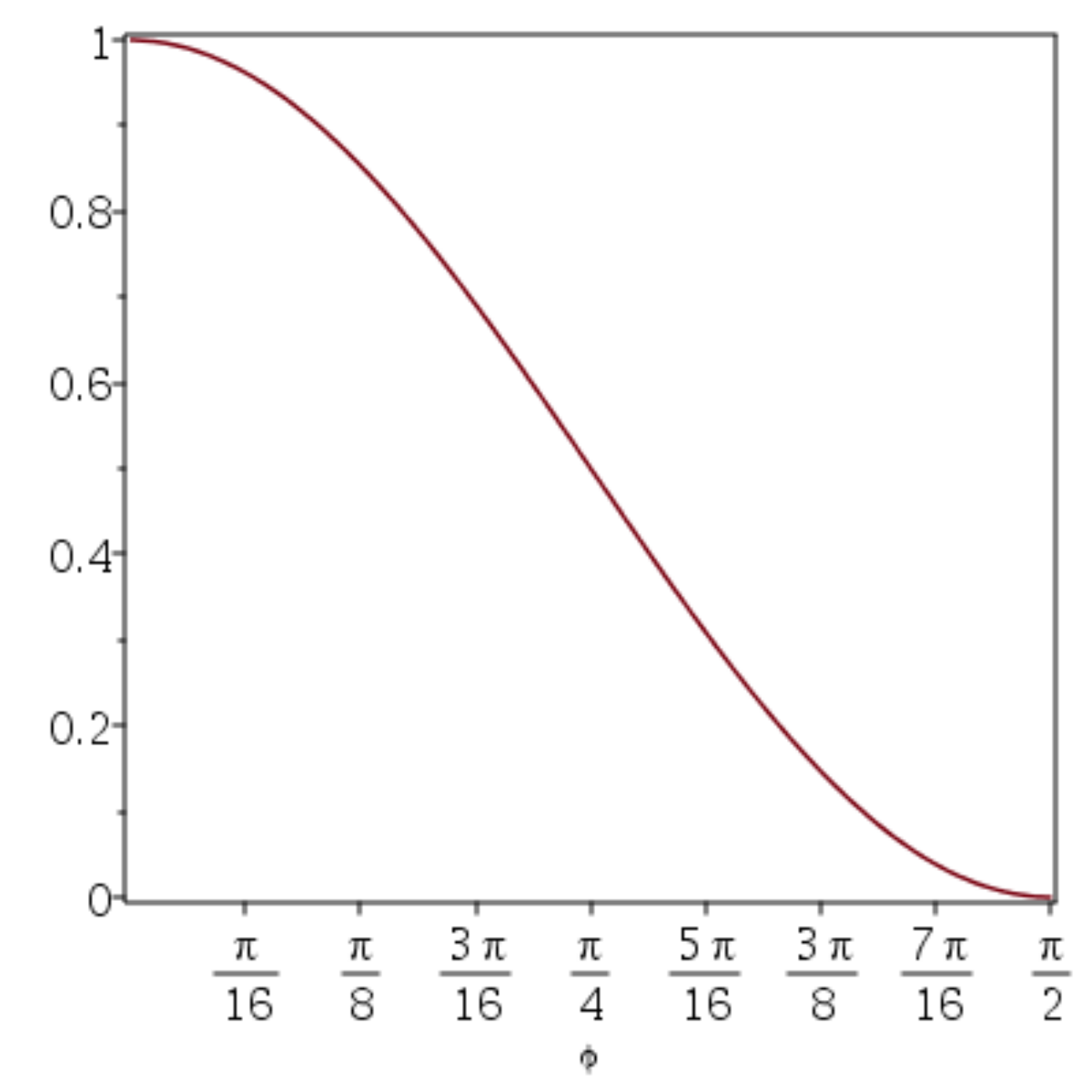}
\caption{Quantifying $\delta\phi$ in terms of $\cos^{-1} \sqrt {\tan \theta_q\over \tan\theta_p}$ for two-step descent towards the equator.}
\label{F:delta-phi}
\end{center}
\end{figure}

{\bf Application to the Kochen-Specker theorem:}\\
Consider any point $p_0$ such that $v(p)=1$ and rotate to put it at the north or south pole.
Then by hypothesis $v(equator)=0$ for any point on the equator.

Now consider any other point $p$ such that $v(p)=0$ and $p$ is not on the equator.
Consider the descent circle $C(p)$; we have $v(p)=0$ by hypothesis, and $v(s)=0$ at the perpendicular point $s$ with $\vec s=(0,-\sin\phi_0,\cos\phi_0)$ where $C(p)$ crosses the equator. Therefore $v(C(p))\equiv 0$ everywhere on this descent circle.

But in particular this implies that $v(r)=0$.
Thence 
$v(C(r))\equiv 0$ everywhere on this descent circle.
Thence $v(q)=0$. This means we have proved:

{\bf Lemma:} If $v(pole)=1$ and $v(p)=0$ then also $v(q)=0$ for $q$ any point on the same line of longitude (same meridian) as $p$ that is closer to the equator than $p$. 
\hfill $\Box$

\begin{figure}[!htbp]
\begin{center}
\includegraphics[scale=2.5]{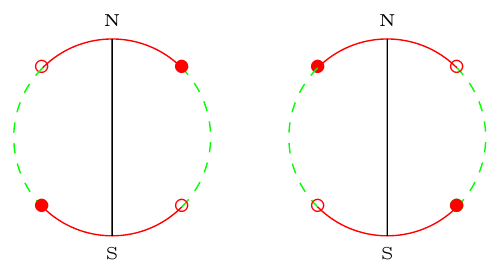}
\caption{Assuming $v(poles)=1$, as a consequence of the two-step descent argument \emph{any} meridian can be put into one of these two forms for \emph{some} value of $\theta_*$.}
\label{F:piece-wise-connected}
\end{center}
\end{figure}

Consequently, for any line of longitude for which $v(poles)=1$,  we see that $v^{-1}(0)$ is path connected. 
Specifically this implies that  $\exists \; \pi/2\leq \theta_*\leq 0$ such that either
\begin{equation}
v(\theta) =  \left\{  \begin{array}{cl} 
\vphantom{\Big |} 1 & \hbox{ for } {\pi\over2}\geq \theta \geq \theta_*;\\
\vphantom{\Big |} 0 & \hbox{ for }  \theta_* > \theta \geq \theta_* - {\pi\over2};\\
\vphantom{\Big |}  1 & \hbox{ for } \theta_* - {\pi\over2} > \theta \geq -{\pi\over2};\\
   \end{array}\right.
\end{equation}
or
\begin{equation}
v(\theta) =  \left\{  \begin{array}{cl} 
\vphantom{\Big |} 1 & \hbox{ for } {\pi\over2}\geq \theta > \theta_*;\\
\vphantom{\Big |} 0 & \hbox{ for }  \theta_* \geq \theta > \theta_* - {\pi\over2};\\
\vphantom{\Big |}  1 & \hbox{ for } \theta_* - {\pi\over2} \geq \theta \leq -{\pi\over2}.\\
   \end{array}\right.
\end{equation}
(See figure \ref{F:piece-wise-connected}.)

Now pick any specific line of longitude, 
by interchanging the north and south poles we can without loss of generality assert
\begin{equation}
v(\theta) =  \left\{  \begin{array}{cl} 
\vphantom{\Big |} 1 & \hbox{ for } {\pi\over2}\geq \theta \geq \theta_*;\\
\vphantom{\Big |} 0 & \hbox{ for }  \theta_* > \theta \geq \theta_* - {\pi\over2};\\
\vphantom{\Big |}  1 & \hbox{ for } \theta_* - {\pi\over2} > \theta \geq -{\pi\over2}.\\
   \end{array}\right.
\end{equation}

Now rotate the sphere $S^2$ around the polar axis so that the line of longitude we have chosen lies on the zero meridian $\phi_*=0$ (the prime meridian). 

\begin{figure}[!htbp]
\begin{center}
\includegraphics[scale=3.0]{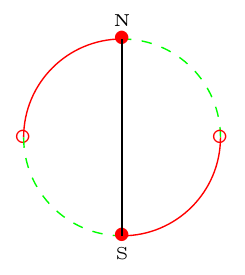}
\caption{Assuming $v(poles)=1$,  after suitable rotations the prime meridian can always be put into this standardized form.}
\label{F:standardized}
\end{center}
\end{figure}

Then furthermore rotate the sphere $S^2$ around the axis perpendicular to the zero meridian so that point $p_*= (\sin\theta_*, 0, \cos\theta_*)$ is moved to the north pole. That is:

{\bf Lemma:}
 Without any loss of generality we can choose the zero meridian to satisfy
\begin{equation}
v(\theta) =  \left\{  \begin{array}{cl} 
\vphantom{\Big |} 1 & \hbox{ for } \theta = {\pi\over2};\\
\vphantom{\Big |} 0 & \hbox{ for }  {\pi\over2} > \theta \geq 0;\\
\vphantom{\Big |}  1 & \hbox{ for }  0> \theta \geq -{\pi\over2}. \\
\end{array}\right.
\label{E:key}
\end{equation}
(See figure~\ref{F:standardized}.)

This will now quickly lead to a contradiction. \hfill $\Box$

First consider all the descent great circles $C(p)$ based on this particular choice of zero meridian. 
These descent great circles will (in the northern hemisphere) sweep out the entire half-hemisphere $\phi\in(-\pi/2,+\pi/2)$ and $\theta\in(\pi/2,0)$.
Similarly, in the southern hemisphere these decent circles will in turn sweep out  the complementary half-hemisphere $\phi\in(+\pi/2,\pi] \cup [-\pi,-\pi/2)$ and $\theta\in(0,-\pi/2)$.
But, following previous arguments,  since $v(\theta)=0$ at the apex of all these descent great circles, $v(C(p))=0$ for all these descent great circles. 
That is:

{\bf Lemma:}
Without loss of generality we have chosen the zero meridian such that (except possibly at the poles themselves) 
\begin{equation}
v(\theta>0,|\phi|<\pi/2) =0; \qquad\hbox{and} \qquad v(\theta<0,|\phi|<\pi/2) =1; 
\label{E:zero1}
\end{equation}
\begin{equation}
v(\theta<0,|\phi|>\pi/2) =0; \qquad\hbox{and} \qquad v(\theta>0,|\phi|>\pi/2) =1.
\label{E:zero2}
\end{equation}

\begin{figure}[!htbp]
\begin{center}
\includegraphics[scale=1.0]{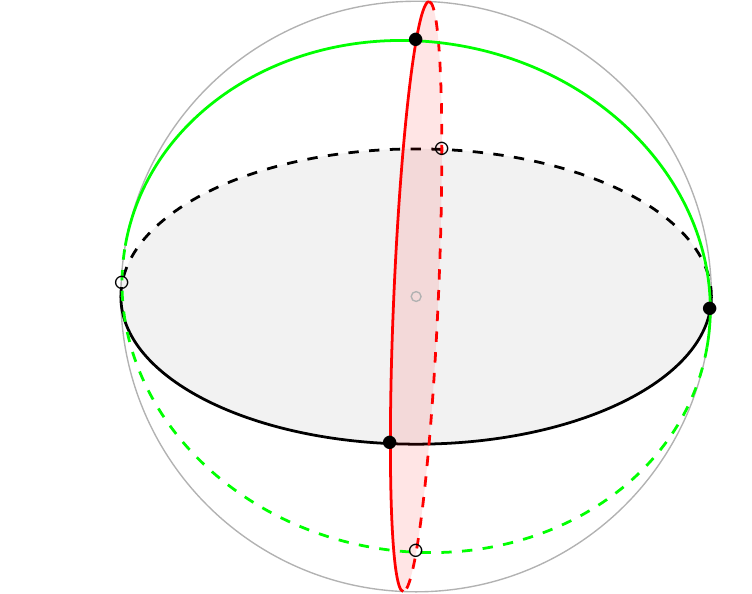}
\caption{Let green denote the prime meridian at $\phi=0$, black the equator at $\theta=0$, and red the meridians at $\phi=\pm\pi/2$. The equator and red meridians split the sphere into four segments, with two of these segments having valuation zero, and two segments having valuation unity.}
\label{F:segments}
\end{center}
\end{figure}

Thus the valuation $v(\cdot)$ is 50\%--50\% zero-one over the entire 2-sphere $S^2$. \hfill $\Box$

Completing the inconsistency argument can now be done in many ways (in fact, an infinite number of ways).
Consider any meridian with $\phi_*\neq 0$ and $|\phi_*| <\pi/2$. On the one hand this meridian will also have the same valuation, equation~(\ref{E:key}), as the zero meridian.
On the other hand  by considering the descent great circles based on this new meridian we have
\begin{equation}
v(\theta>0,|\phi-\phi_*|<\pi/2) =0; \qquad\hbox{and} \qquad v(\theta>0,|\phi-\phi_*|<\pi/2) =1; 
\end{equation}
\begin{equation}
v(\theta<0,|\phi-\phi_*|>\pi/2) =0; \qquad\hbox{and} \qquad v(\theta<0,|\phi-\phi_*|>\pi/2) =1.
\end{equation}
But this is incompatible with the behaviour based on the zero meridian, equations~(\ref{E:zero1}) and (\ref{E:zero2}), so we have a contradiction.

This completes the proof of Kochen--Specker in three dimensions. We feel that this is a nice simple proof of Kochen--Specker that does not rely on finding explicit bases for the Hilbert space --- it also seems to us  to be considerably simpler than the other geometric or colouring arguments.

\subsection{$d = 4$ and higher dimensions}

What happens in a $d>3$-dimensional Hilbert space? 

The 3-dimensional logic carries over with utterly minimal modifications. 
\begin{itemize}
\item 
In $d=4$ one needs to study the unit 3-sphere $S^3$. Pick any point $n$ on $S^3$ such that $v(n)=0$. This can always be done. Then consider the 2-sphere perpendicular to chosen point $n$.
On that 2-sphere the 4-dimensional Kochen--Specker theorem will reduce to the 3-dimensional Kochen--Specker theorem, which we have already established. So nothing more need be done.
\item
In $d\geq 4$ dimensions one needs to study the unit $(d-1)$-sphere $S^{d-1}$. Pick any $d-3$ mutually-orthogonal points $n_i$ on $S^3$ such that $v(n_i)=0$.  If this cannot be done then the existence of the claimed valuation $v(\cdot)$  already fails at this elementary level so that the $d$-dimensional Kochen--Specker theorem is established; so without loss of generality we can assume this can be done.
Then consider the 2-sphere perpendicular to all the $n_i$.
On that 2-sphere the $d$-dimensional Kochen--Specker theorem will reduce to the 3-dimensional Kochen--Specker theorem, which we have already established. So nothing more need be done.
\end{itemize}
It is interesting to note that 3-dimensions is the key part of the theorem; in 1 and 2 dimensions related results are trivial. In 4 or more dimensions the Kochen--Specker theorem follows immediately from the 3-dimensional result. 

\section{Relation to Gleason's theorem}

The  Kochen--Specker theorem is  more basic and fundamental than the equally well-known Gleason's theorem~\cite{Gleason,Cooke,without,Del}. 
Indeed, if one assumes Gleason's theorem then the Kochen--Specker theorem is trivial. 
The point is that once one asserts that the valuation $v(\cdot)$ is inherited from a density matrix $v(n) = \langle n| \rho| n \rangle$, 
then one knows that the valuation is continuous. But no function from the connected space $S^n$ to the discrete set $\{0,1\}$ (with its implied discrete topology) can possibly be continuous. 

\section{Discussion}

It has been exactly 50 years since the groundbreaking publication of the theorem due to Kochen and Specker~\cite{Kochen:1967}.  Along with Bell's inequality, their discovery represents one of the two major no-go theorems in the foundations of quantum theory~\cite{Kochen:1967,Bell:1966}.  The main implication of the result is that quantum theory fails to allow a non-contextual hidden variable model.  More precisely, it states that it is impossible for the predictions of quantum mechanics to be in line with measurement outcomes which are pre-determined in a non-contextual manner.  Hence this would rule out a large class of hidden variable models that might otherwise seem at first sight to be intuitive representations of  the physical world.

\enlargethispage{20pt}
Furthermore, there is increasing support that the notion of contextuality captures the essential difference between the quantum and classical world.  A specific case of this would be in the recent evidence that contextuality may be the primary reason for the speedup of universal quantum computation.  This has been shown through `magic' state injection~\cite{compute}.
Recent work has also connected  contextuality with non-locality~\cite{contextuality}.  With non-locality almost ubiquitous in the field of quantum information, this connection only serves to increase the impact and significance  of research into this lesser known property.  In addition to theoretical results, recent experimental tests to demonstrate contextuality include a superconducting qutrit implementation~\cite{contextuality2}, as well as a photonic implementation~\cite{contextuality3}.

Simpler proofs of the Kochen--Specker theorem were found in years following the original paper~\cite{Kochen:1967}, but these mainly involved the mathematically ``elementary'' (but technically subtle and demanding) machinery of setting up a large number of carefully chosen projection operators or increasing the minimum dimension from three to four or even eight~\cite{Peres,Kernaghan,Cabello1,Cabello2,Peres2}.  

In this article, we have presented a geometric approach where one constructs and exploits the properties of great circles on a $n$-sphere.  This has the power to significantly simplify the argument, while maintaining the validity of the theorem for a minimum dimension of three.   We hope that with this simplification, further work might extend the argument to develop newer results in the foundations of quantum physics.

\enlargethispage{20pt}
\acknowledgments{
\vspace{-10pt}
DR is indirectly supported by the Marsden fund, \\
administered by the Royal Society of New~Zealand.
\\
MV is directly supported by the Marsden fund, \\
administered by the Royal Society of New~Zealand.}

\clearpage
\bigskip
\hrule
\vspace{-10pt}
\section*{Background resources}
\medskip
\hrule
\medskip

For a general introduction to the Kocher--Specker theorem, see:
\begin{itemize}
\item
 \url{https://en.wikipedia.org/wiki/Kochen--Specker_theorem}
\item
\url{https://plato.stanford.edu/entries/kochen-specker/}
\end{itemize}

\bigskip
\hrule
\vspace{-10pt}
 
\end{document}